\documentclass[twocolumn,twoside,slac_two]{revtex4}
\usepackage{graphicx}
\usepackage{fancyhdr}
\begin{document}

\title{Polarization signatures of strong gravity in black-hole accretion discs}

\author{Vladim\'{\i}r Karas}
\affiliation{Astronomical Institute, Academy of Sciences, Prague, Czech Republic}
\affiliation{Charles University, Faculty of Mathematics and Physics, Prague, Czech Republic}

\author{Michal Dov\v{c}iak}
\affiliation{Astronomical Institute, Academy of Sciences, Prague, Czech Republic}

\author{Giorgio Matt}
\affiliation{Dipartimento di Fisica, Universit\`a degli Studi ``Roma Tre'', Rome, Italy}

\begin{abstract}
We discuss the effects of strong gravity on the polarization properties of a
black hole accretion disc. The intrinsic polarization is computed taking
into account light scattered on the disc surface and using different
approximations. The gravitational field of a black hole influences the
Stokes parameters of reflected radiation propagating to a distant
observer. The lamp-post model is explored as an example of a specific
geometrical arrangement relevant for AGNs. The degree and the angle of
polarization are computed as functions of the observer inclination
angle, of the inner radius of the disc emitting region, and of other
parameters of the model. The expected polarization should be detectable
by new generation polarimeters.
\end{abstract}
\maketitle

\thispagestyle{fancy}

\section{Introduction}
Accretion discs in central regions of active galactic nuclei are
subject to strong external illumination originating from some kind
of corona and giving rise to specific spectral features in the X-ray
band. In particular, the K-shell lines of iron are found to be prominent
around $6$--$7$~keV. It has been shown that the shape of the intrinsic
spectra must be further modified by the strong gravitational field of
the central mass, and so X-ray spectroscopy could allow us to explore
the innermost regions of accretion flows near supermassive black holes
\cite{fabian00,reynolds03}.
Similar mechanisms operate also in some Galactic black-hole candidates.

Recent {\it XMM-Newton}\/ observations indicate the rather surprising 
result that relativistic iron lines are not as common as previously believed;
see Bianchi et al.\ \cite{bianchi04} and Yaqoob et al.\ \cite{yaqoob03}
fort further references.
This does not necessarily mean that the iron line is not produced in
the innermost regions of accretion discs but the situation is likely to be more
complex than in simple, steady scenarios. Some evidence for the
line emission arising from orbiting spots is present in the
time-resolved spectra of a few AGNs \cite{dovciak04a}. Even
when clearly observed, relativistic lines behave differently than
expected. The best example is the puzzling lack of correlation  between
the line and continuum emission in MCG--6-30-15 \citep{fabian02},
unexpected because the very broad line profile clearly indicates that
the line originates in the innermost regions of the accretion disc,
hence very close to the illuminating source. Miniutti et al.\ \cite{miniutti03}
have proposed a solution to this problem in terms of an illuminating
source moving along the black-hole rotation axis or very close to it.

Polarimetric studies along with time-resolved spectroscopy could provide
additional information about accretion discs in the strong gravity
regime, and this may be essential to discriminate between different
possible geometries of the source. The idea of using polarimetry to gain
additional information about accreting compact objects is not a new one.

In this context it was proposed by Rees \cite{rees75} that polarized X-rays
are of high relevance. Pozdnyakov  et al.\ \cite{pozdnyakov79} 
studied spectral profiles of iron
X-ray lines that result from multiple Compton scattering.
Various influences affecting polarization (due to magnetic fields,
absorption as well as strong gravity) were examined for black-hole
accretion discs \citep{agol96}.  Temporal variations of polarization were
also discussed, in particular the case of orbiting spots near a black
hole \citep{connors80,bao96}. Furthermore, within general relativity framework
the polarization has been studied by various authors: see
Portsmouth \& Bertschinger \cite{portsmouth04} for a very recent discussion. With
the promise of new polarimetric detectors \citep{costa01},
quantitative examination of specific models becomes timely.

\begin{figure*}[tbh!]
\includegraphics[width=0.328\textwidth]{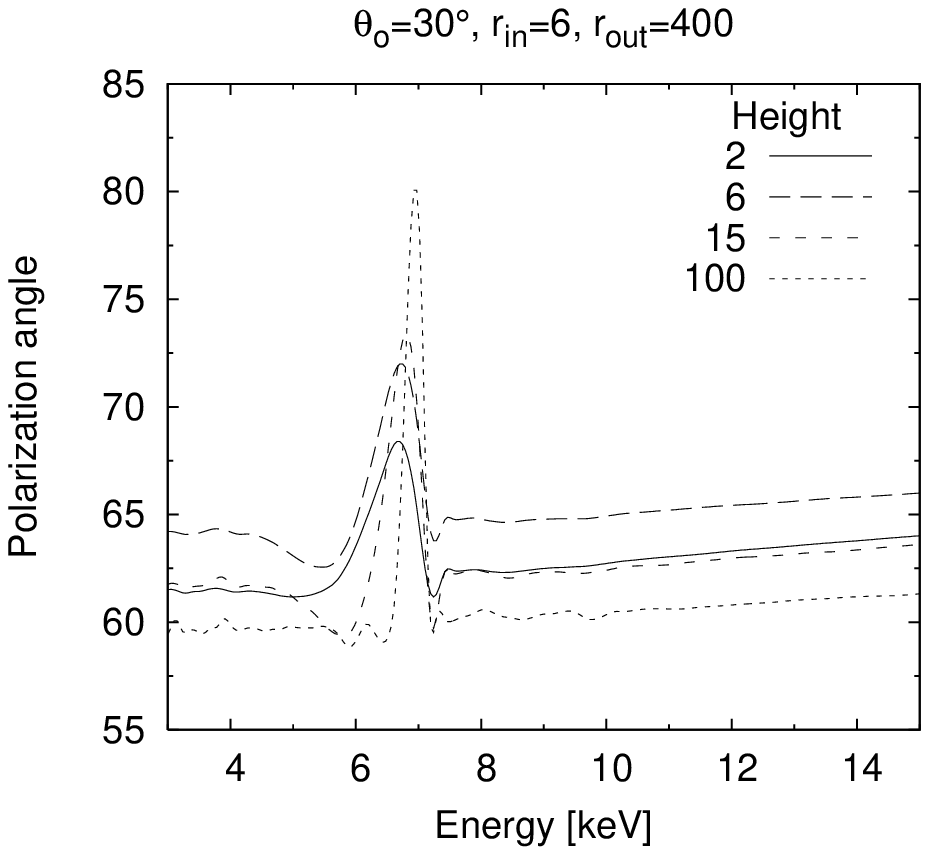}
\hfill
\includegraphics[width=0.328\textwidth]{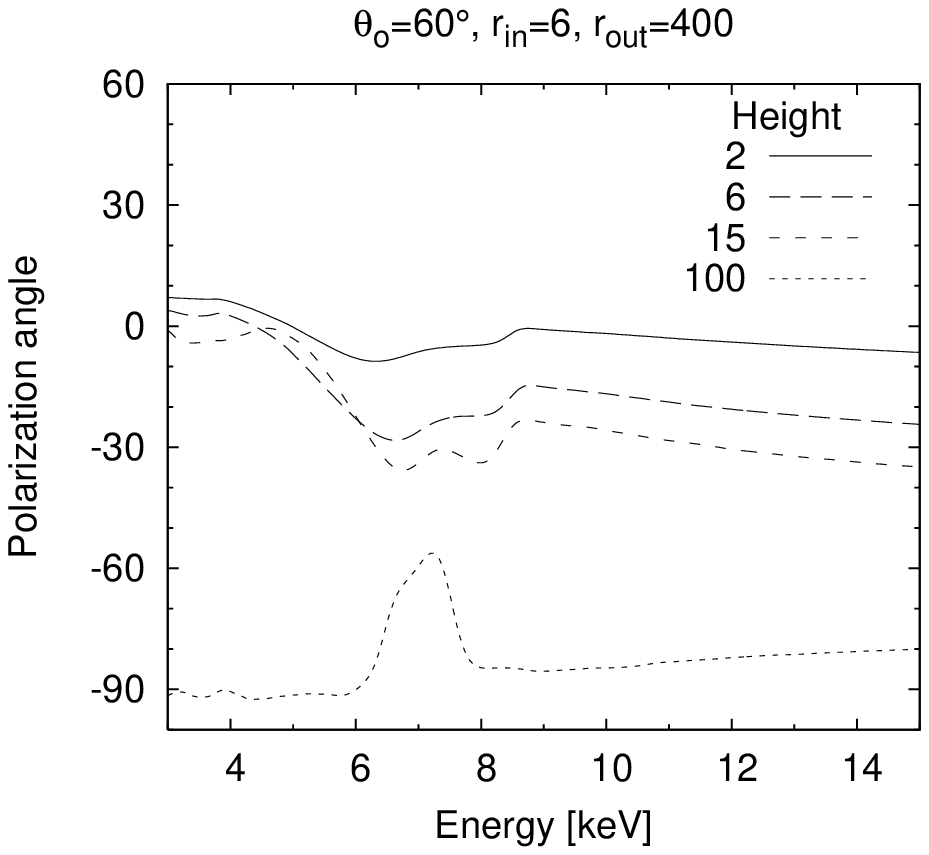}
\hfill
\includegraphics[width=0.328\textwidth]{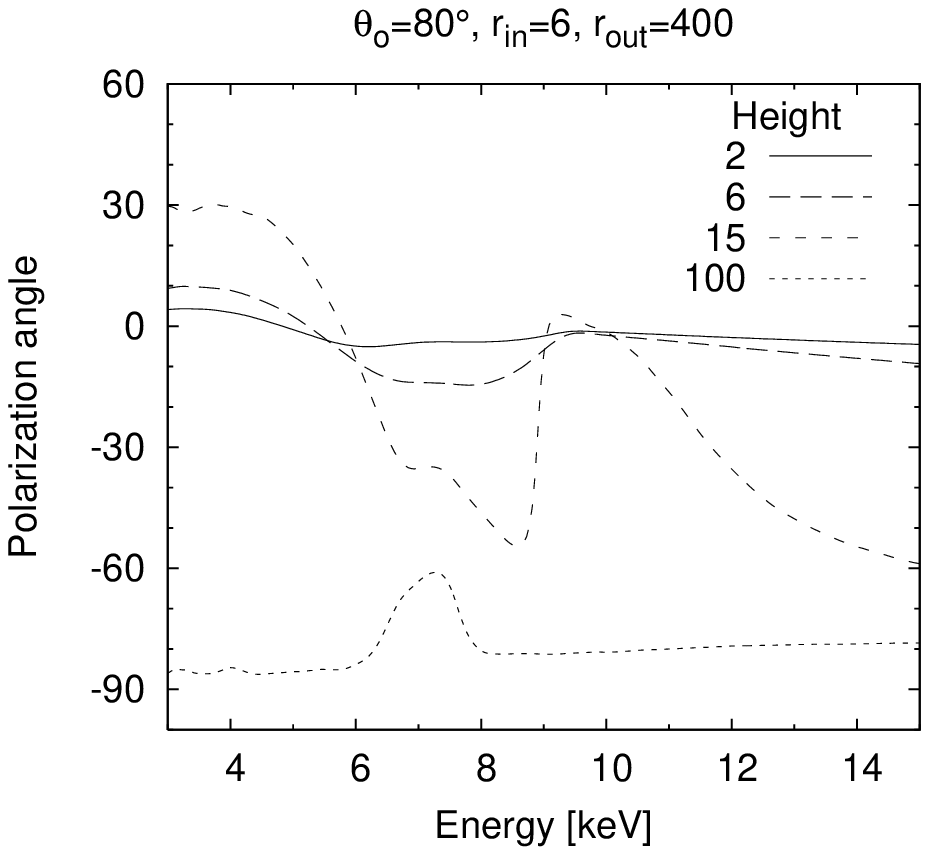}
\includegraphics[width=0.328\textwidth]{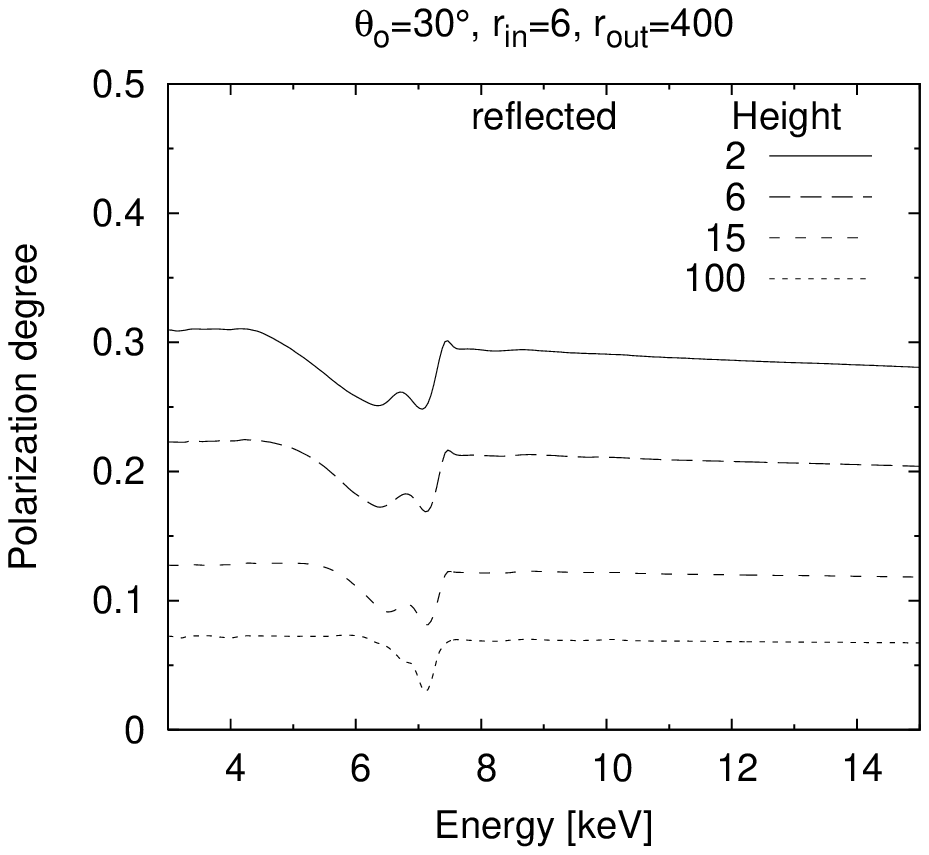}
\hfill
\includegraphics[width=0.328\textwidth]{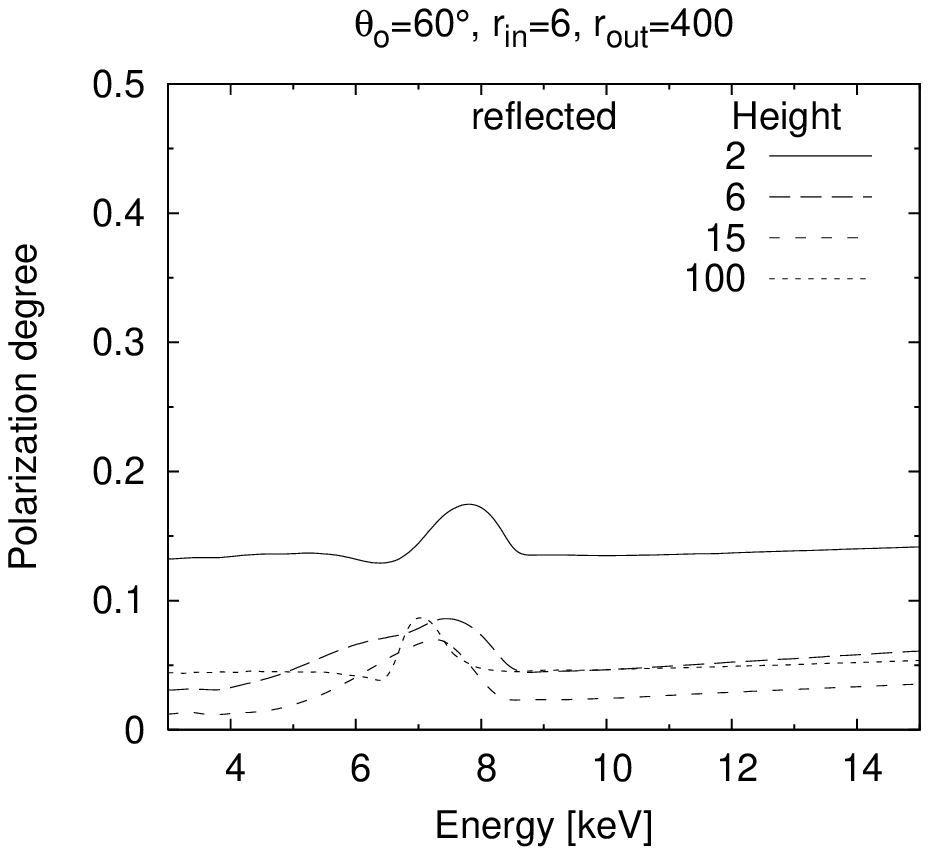}
\hfill
\includegraphics[width=0.328\textwidth]{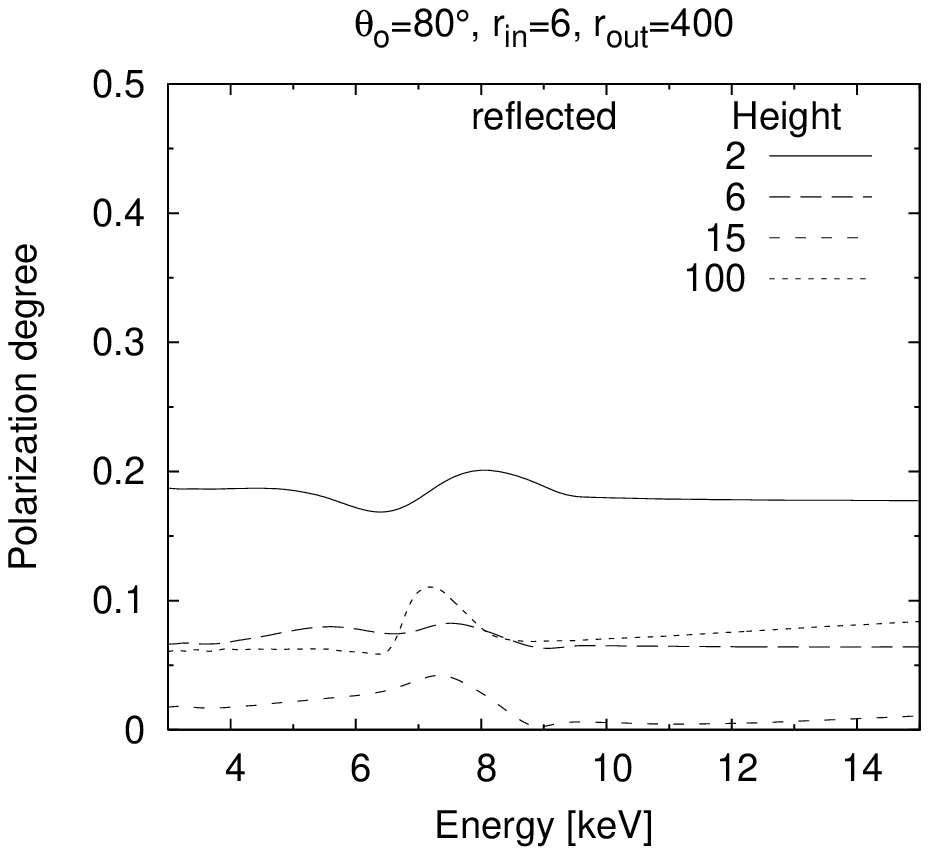}
\includegraphics[width=0.328\textwidth]{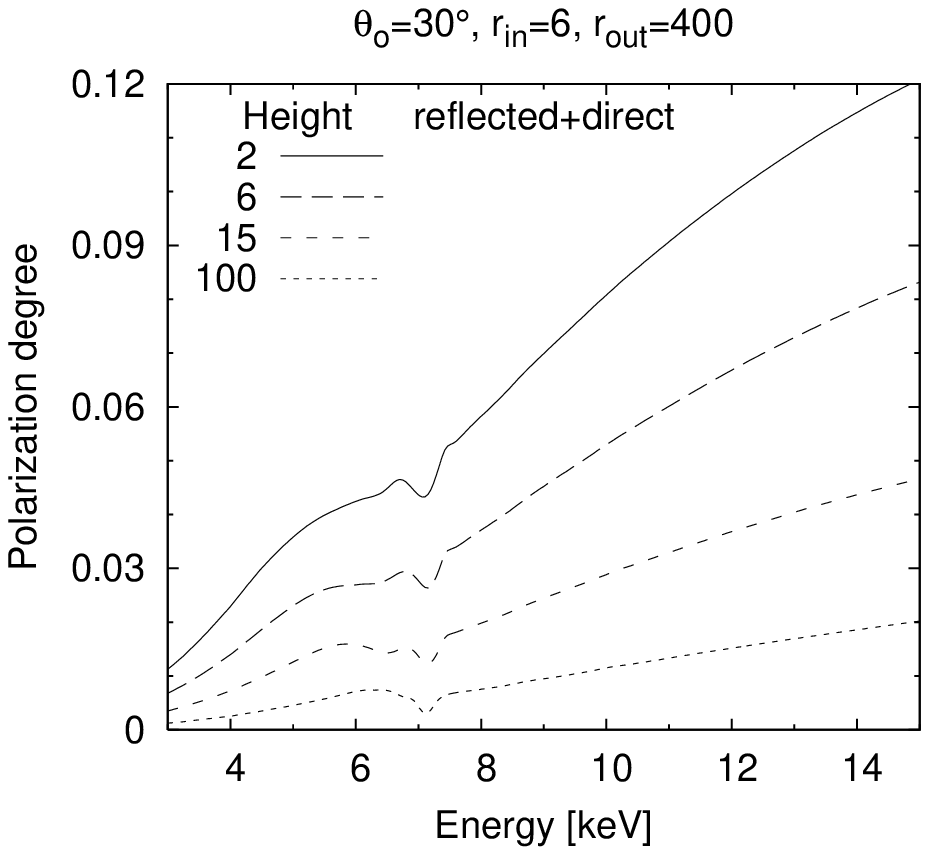}
\hfill
\includegraphics[width=0.328\textwidth]{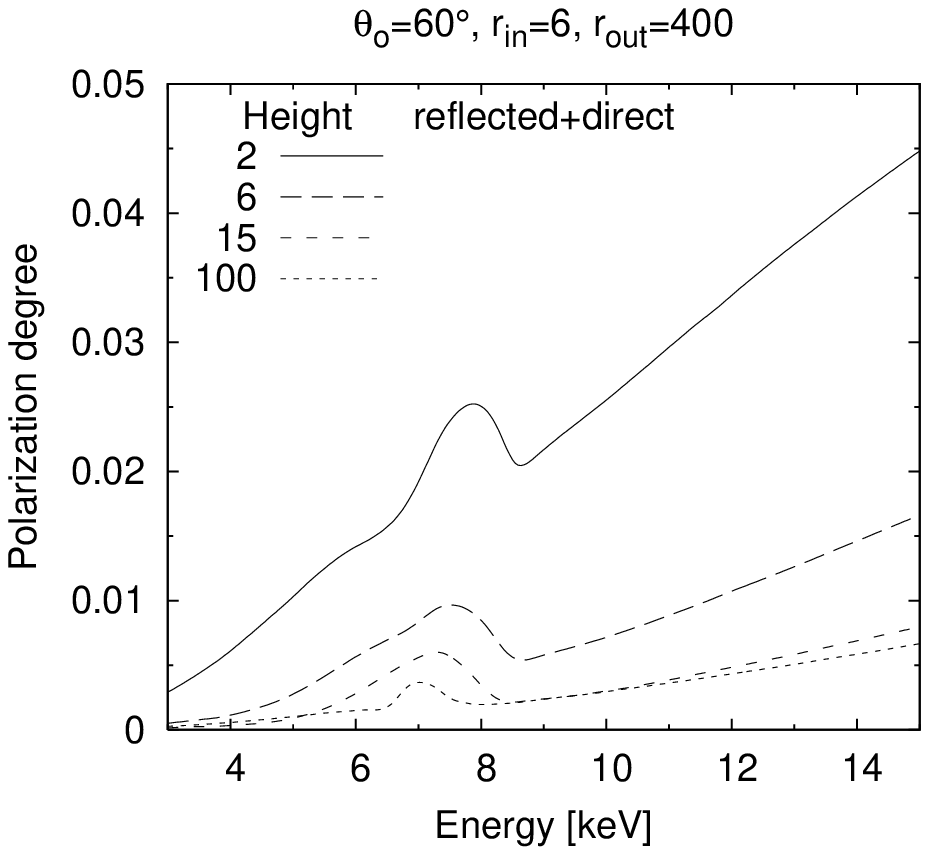}
\hfill
\includegraphics[width=0.328\textwidth]{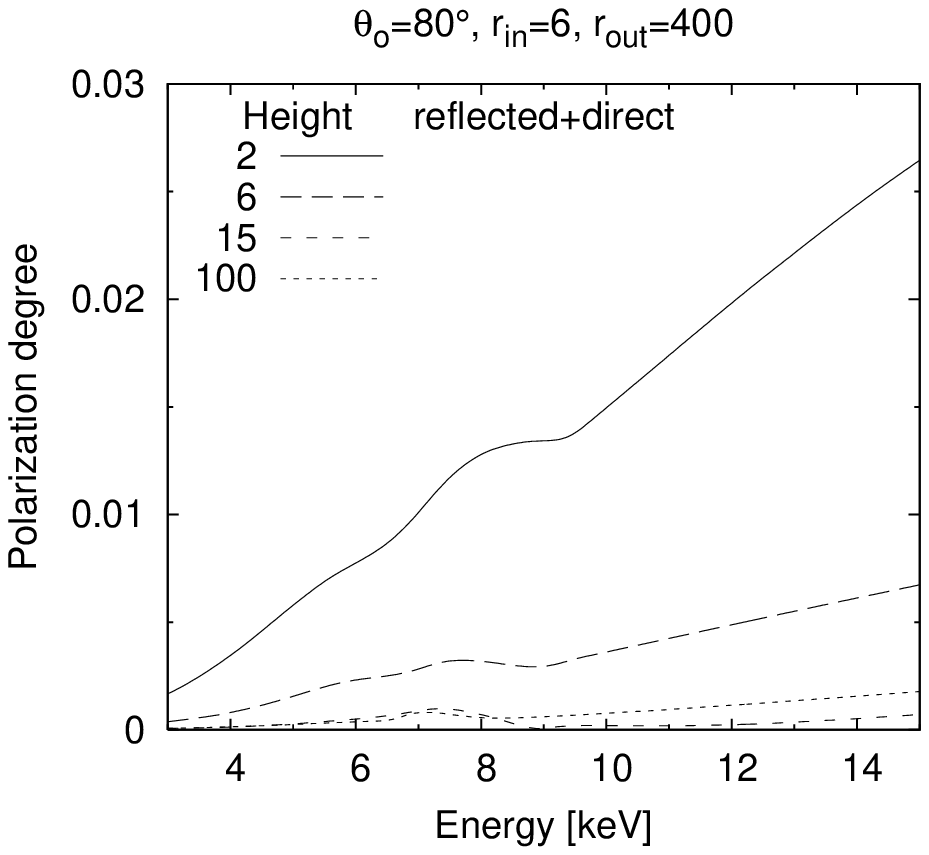}
\caption{The energy dependence of polarization angle (top panels) and polarization
degree (middle panels) due to reflected radiation for different observer's
inclination angles ($\theta_{\rm o}=30^\circ$, $60^\circ$ and $80^\circ$) and for
different heights of the primary source ($h=2$, $6$, $15$ and $100$).
Polarization degree for reflected plus direct radiation is also plotted (bottom
panels). The emission comes from a disc within $r_{\rm{}in}=6$ and
$r_{\rm out}=400$. Isotropic primary radiation with photon index $\Gamma=2$ and
angular momentum of the central black hole $a=0.9987$ were assumed.}
\label{pol}
\end{figure*}

\section{The model and computations}
We have developed a new code for time-dependent analysis of relativistic
spectral features originating in black-hole accretion discs
\cite{dovciak04a,dovciak04c,dovciak04b}. Detailed tracking of time-dependent
spectral features (time-scale of several kiloseconds) would be particularly
relevant for the modelling and interpretation of variable X-ray features.
These have been reported in a growing number of active galactic nuclei
and tentatively interpreted in terms of reflection iron lines due to 
flares and spots \cite{dovciak04b,guainazzi03,turner02,turner04,yaqoob03}.
Narrow fluctuating profiles are often redshifted with respect to
the rest energy and they may provide a powerful tool to measure the
mass of central black holes in AGNs and Galactic black hole candidates
\cite{czerny04,miniutti04}.

Since the reflecting medium has a disc-like geometry, a substantial
amount of linear polarization is expected in the resulting spectrum
because of Compton scattering. Polarization properties of the disc
emission are modified by the photon propagation in a gravitational field,
providing additional information on its structure. Here we
calculate the observed polarization of the reflected radiation assuming
the lamp-post model for the stationary power-law illuminating source
\citep{martocchia96, petrucci97}. We assumed a rotating (Kerr) black hole
as the only source of the gravitational field, having a common symmetry
axis with an accretion disc. The disc was
also assumed to be stationary and we restricted ourselves to the
time-averaged analysis (we assumed processes that vary
at a much slower pace than the light-crossing time at the corresponding
radius). The intrinsic polarization of emerging
light was computed locally, assuming a plane-parallel scattering
layer which was illuminated by light radiated from the primary source.

\begin{figure*}[tb]
\includegraphics[width=0.328\textwidth]{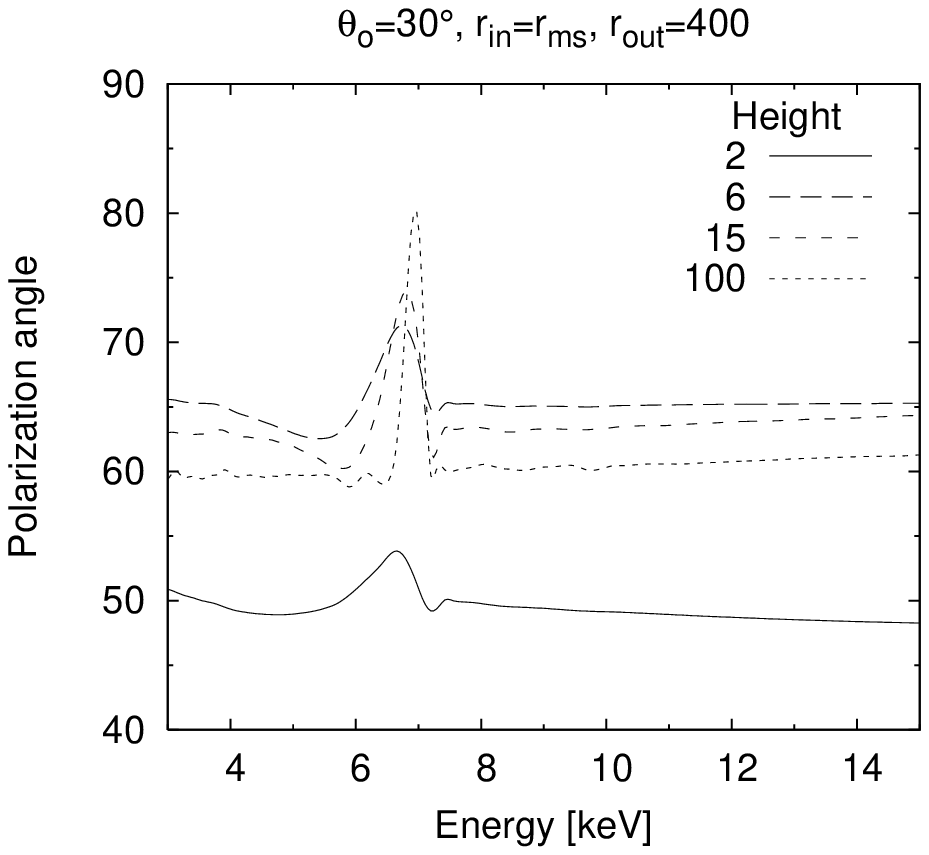}
\hfill
\includegraphics[width=0.328\textwidth]{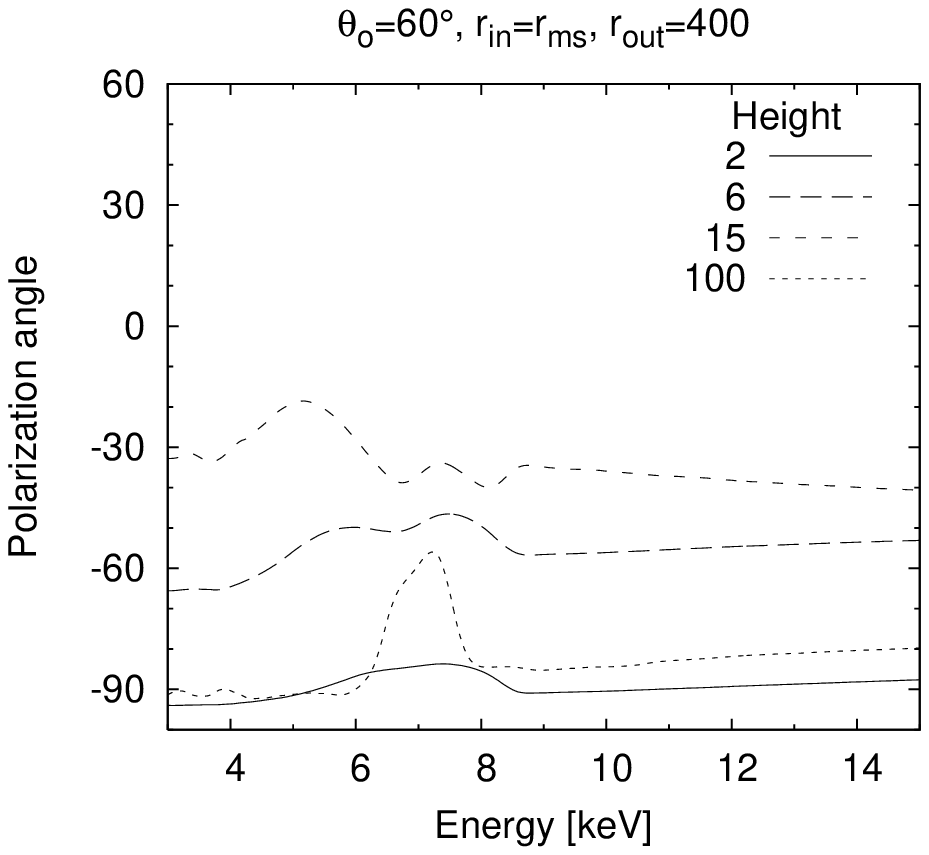}
\hfill
\includegraphics[width=0.328\textwidth]{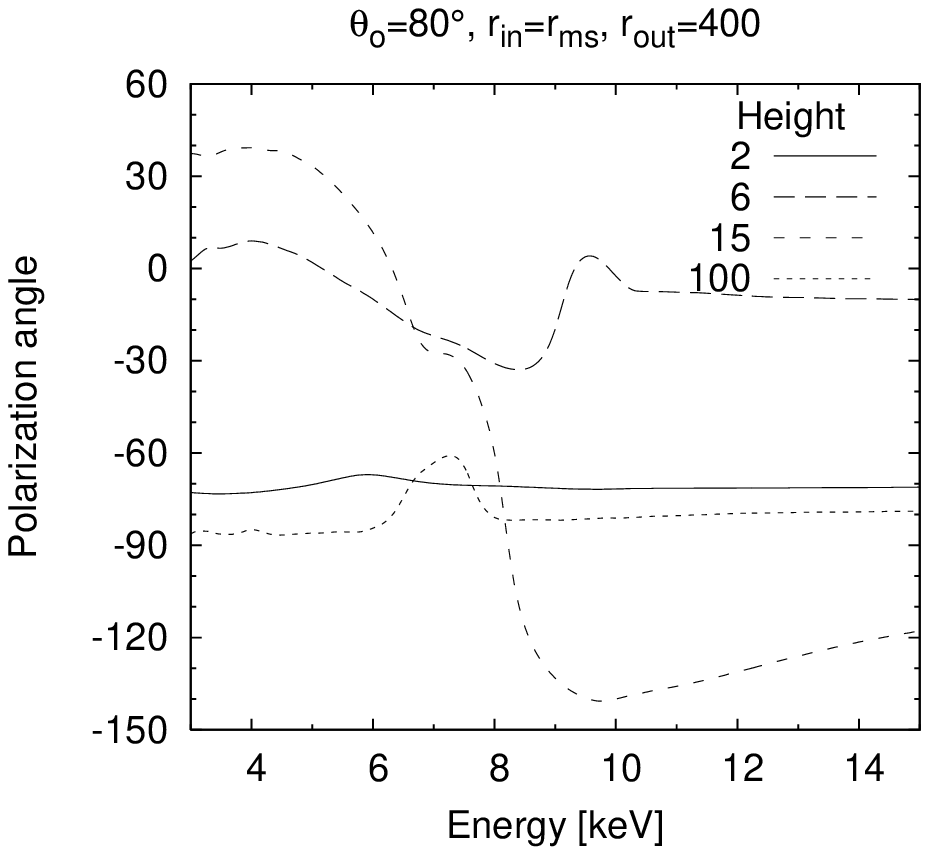}
\includegraphics[width=0.328\textwidth]{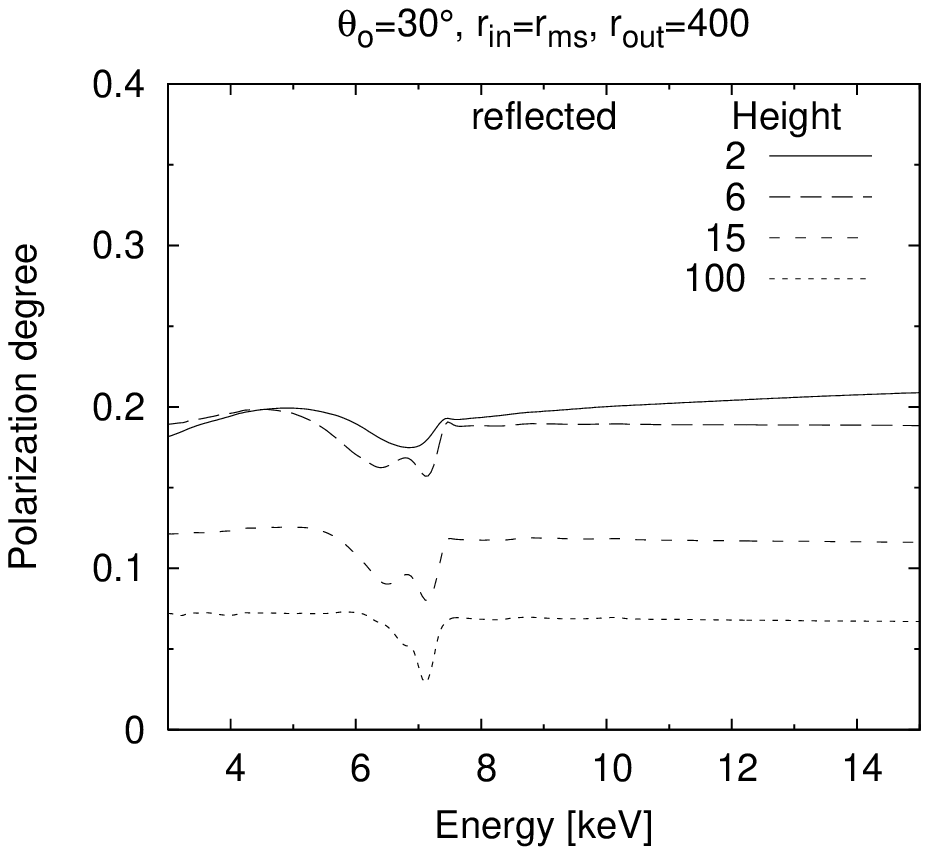}
\hfill
\includegraphics[width=0.328\textwidth]{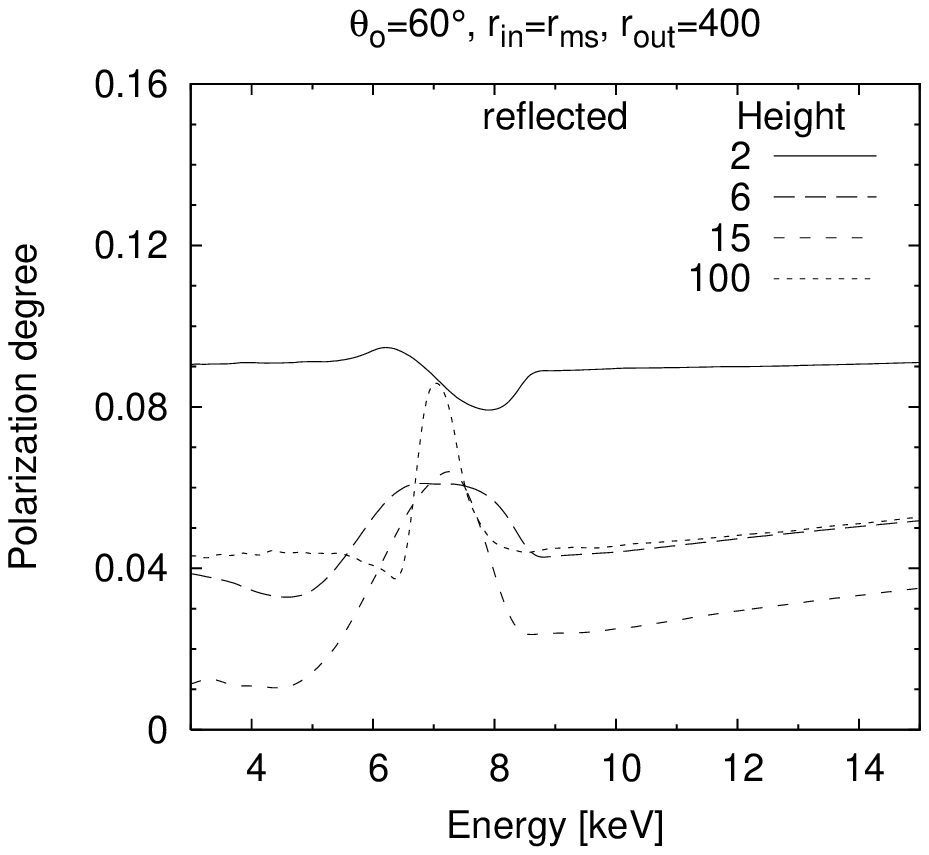}
\hfill
\includegraphics[width=0.328\textwidth]{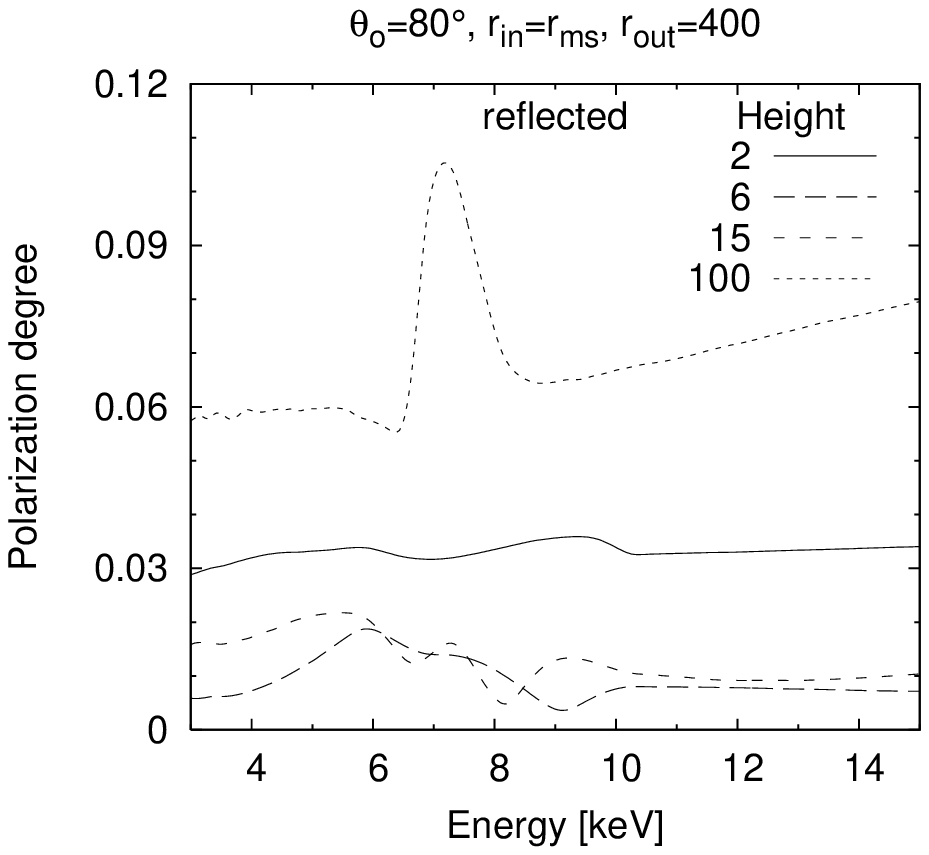}
\includegraphics[width=0.328\textwidth]{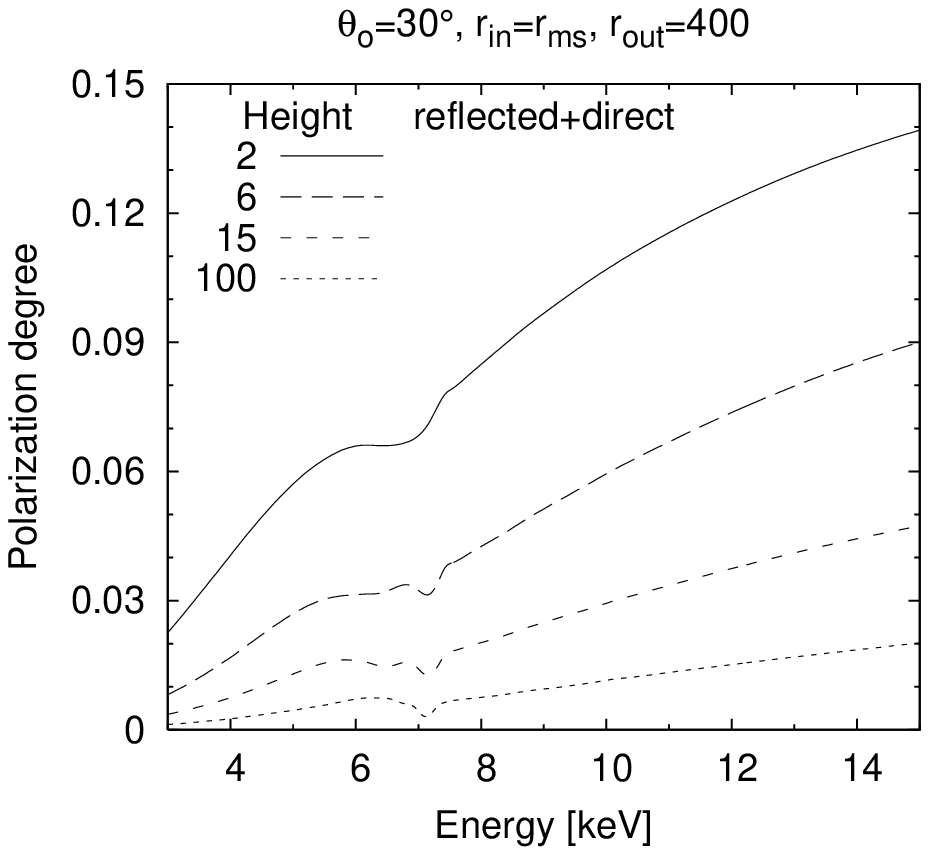}
\hfill
\includegraphics[width=0.328\textwidth]{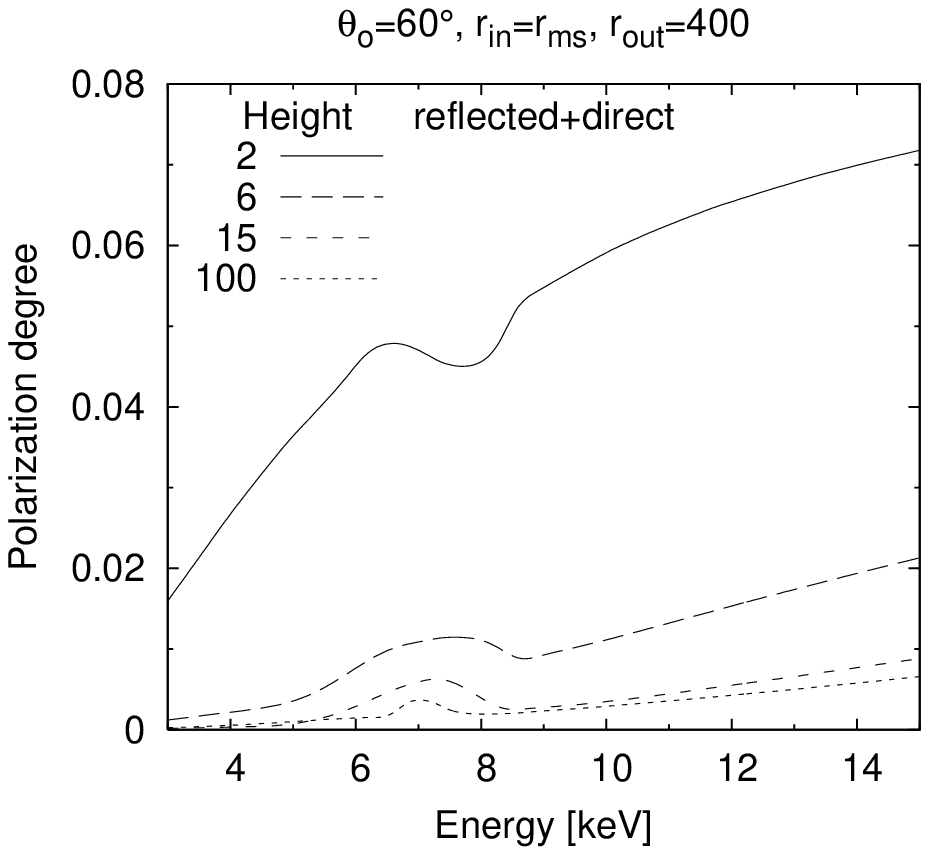}
\hfill
\includegraphics[width=0.328\textwidth]{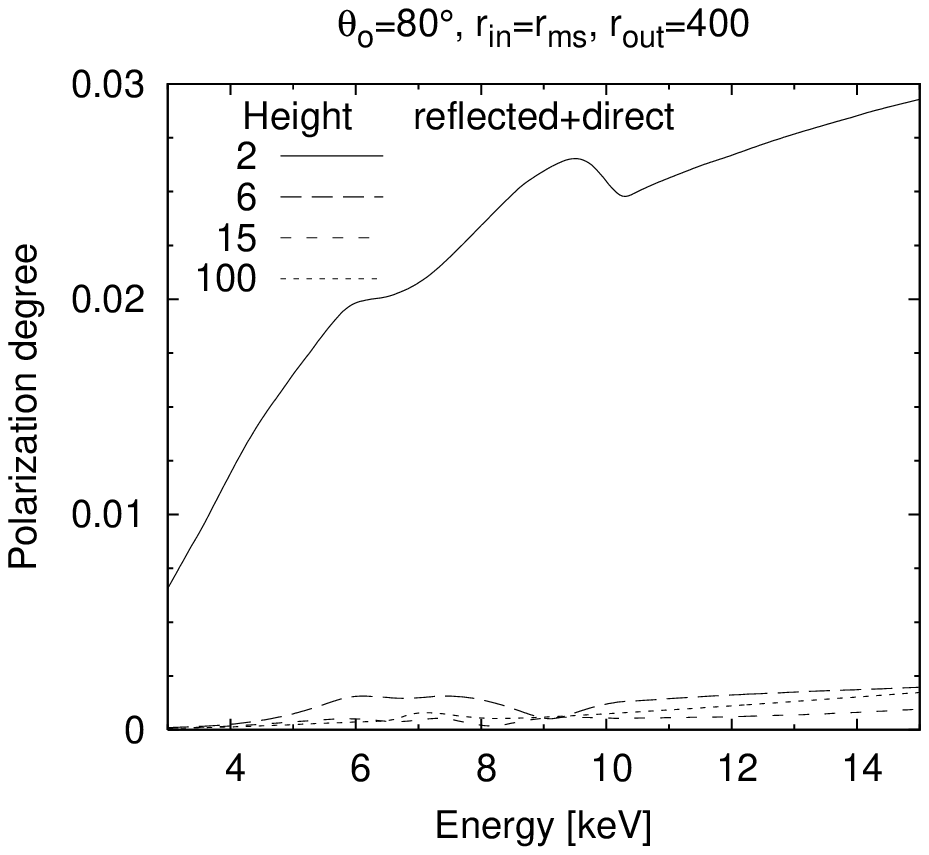}
\caption{Same as in the previous figure but for disc starting at
$r_{\rm in}=1.20\,$.}
\label{pol1}
\end{figure*}

\begin{figure*}
\includegraphics[width=0.45\textwidth]{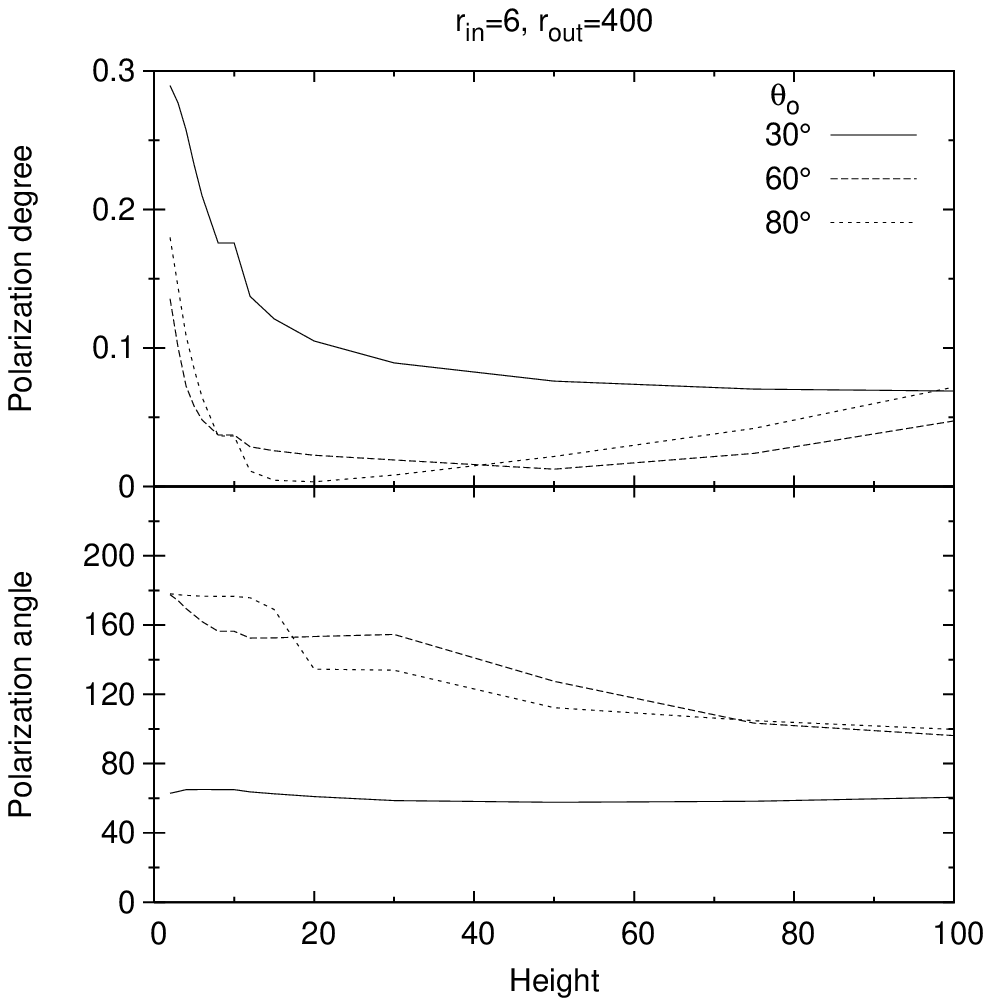}
\hfill
\includegraphics[width=0.45\textwidth]{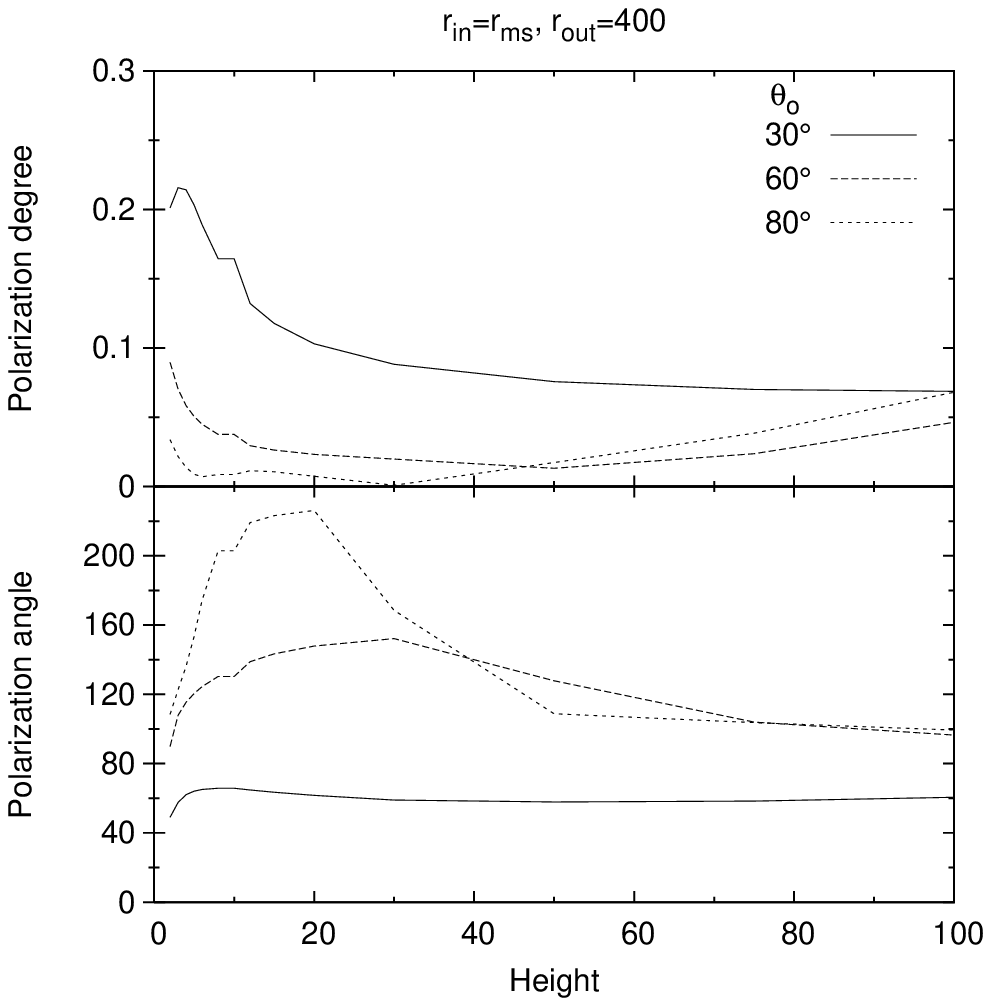}
\caption{Polarization degree and angle due to reflected radiation integrated
over the whole surface of the disc and propagated to
the point of observation. Dependence on height $h$ is plotted.
Left panel: $r_{\rm{}in}=6$; right panel: $r_{\rm{}in}=1.20$. In both
the panels the energy range was assumed $9-12$~keV, the photon index of
incident radiation $\Gamma=2$, the angular momentum $a=0.9987$.}
\label{poldeg}
\end{figure*}

\begin{figure*}
\includegraphics[width=0.45\textwidth]{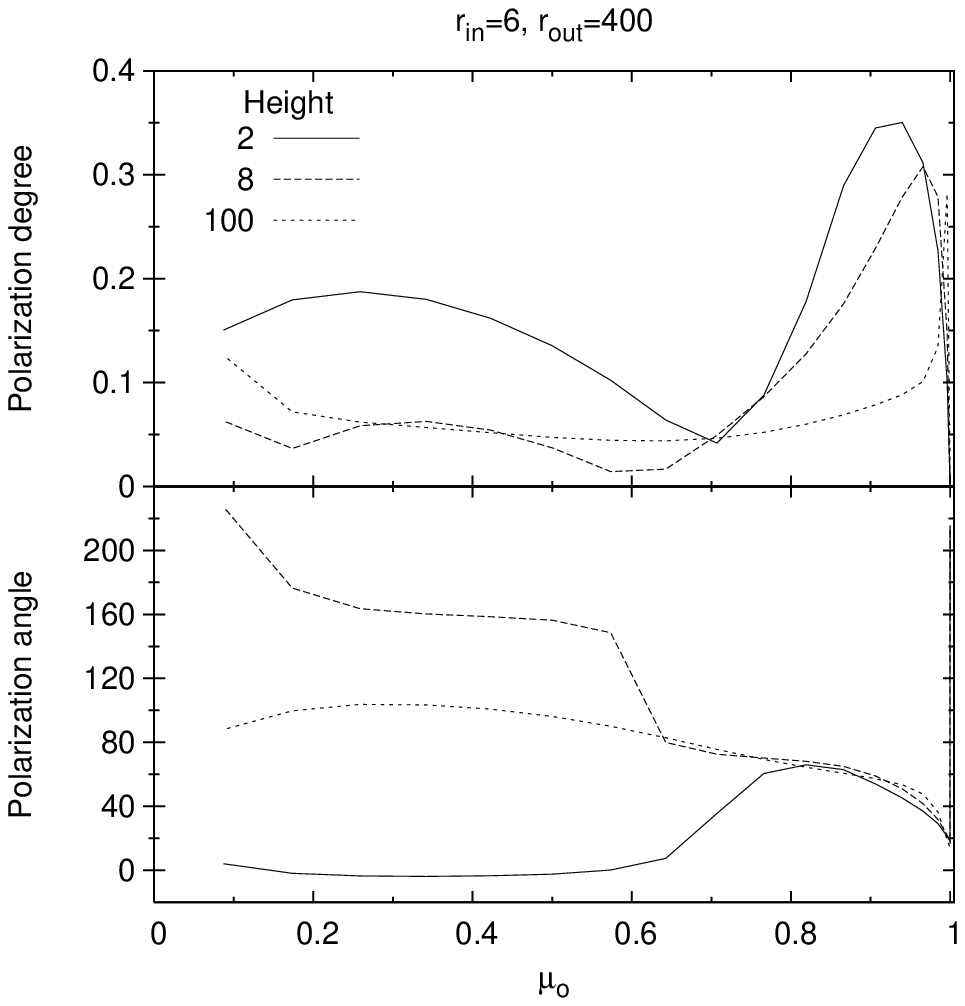}
\hfill
\includegraphics[width=0.45\textwidth]{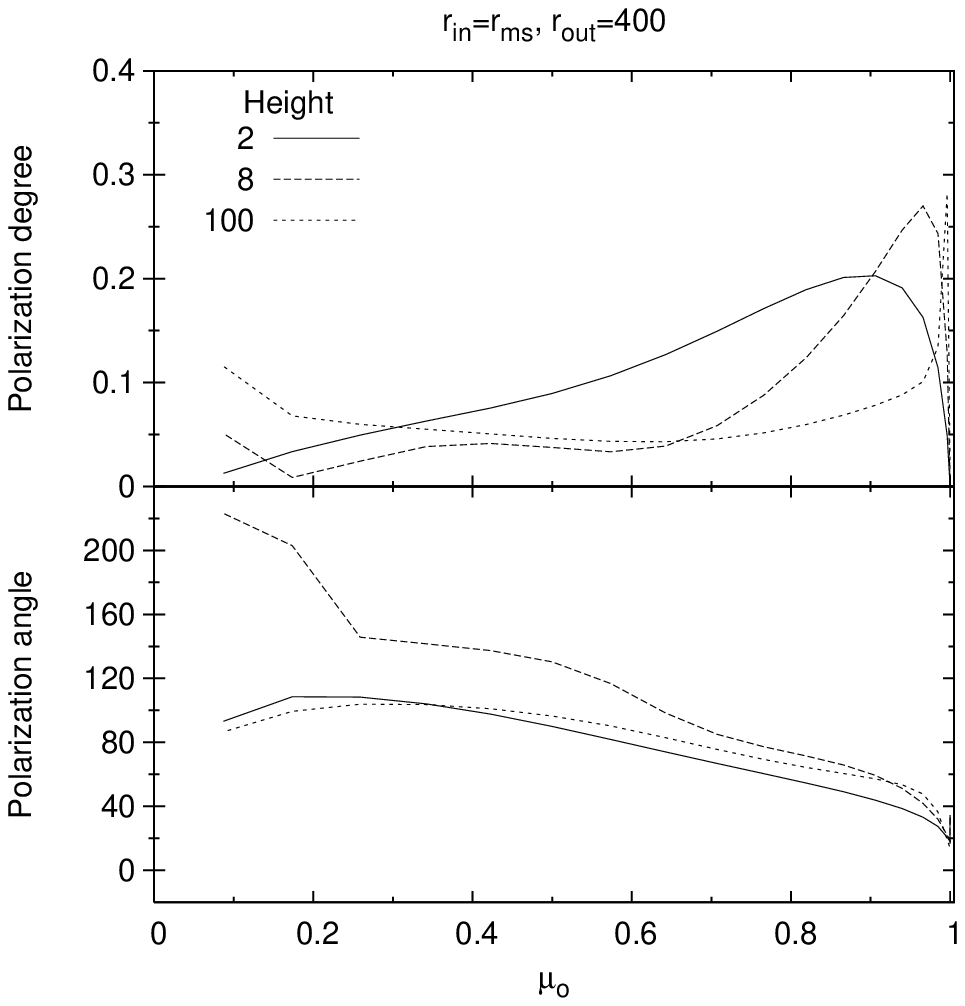}
\caption{Polarization degree and angle as functions of
$\mu_{\rm{}o}$ (cosine of observer inclination, $\mu_{\rm{}o}=0$
corresponds to the edge-on view of the disc). The same model
is shown as in the previous figure.}
\label{polangle1}
\end{figure*}

\begin{figure*}
\includegraphics[width=0.45\textwidth]{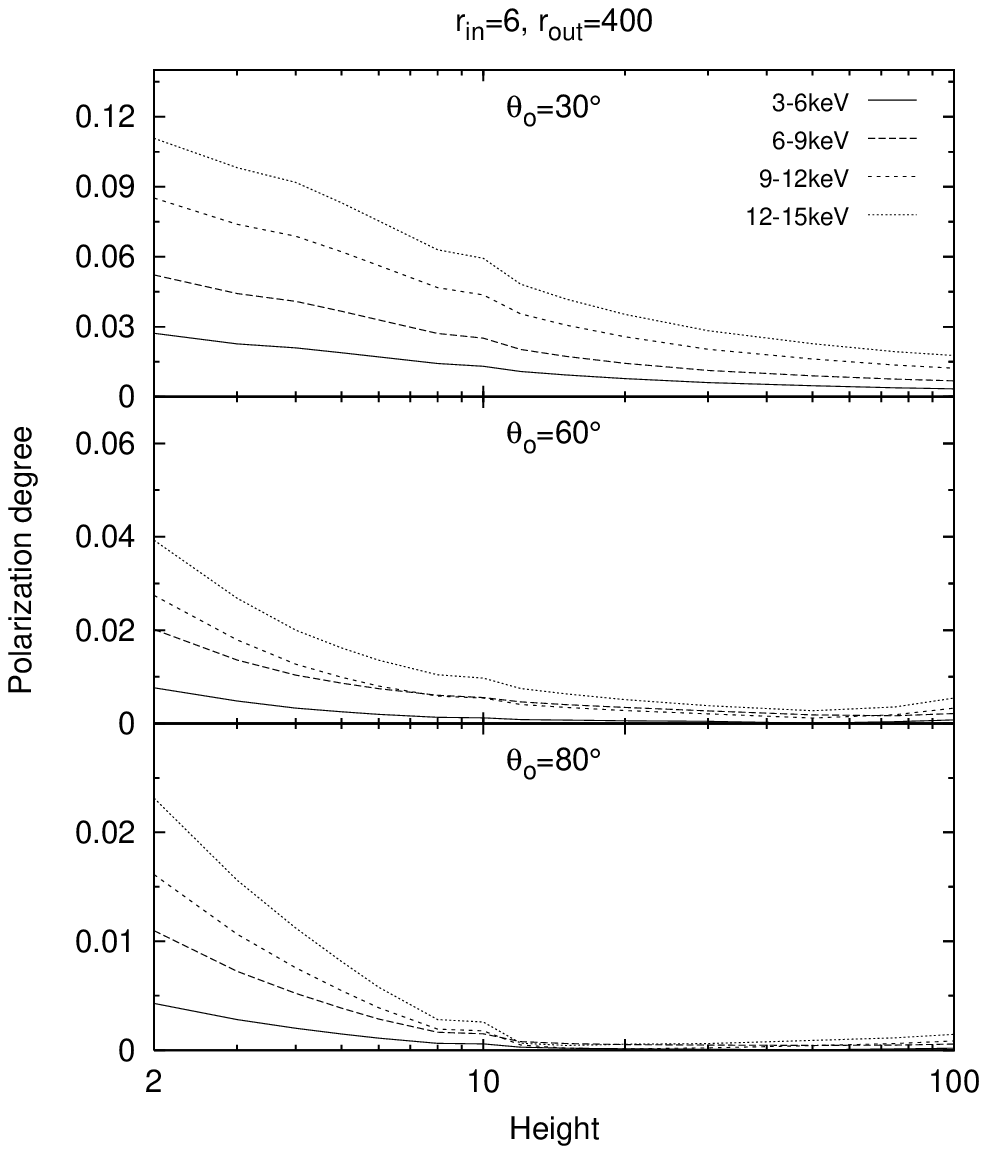}
\hfill
\includegraphics[width=0.45\textwidth]{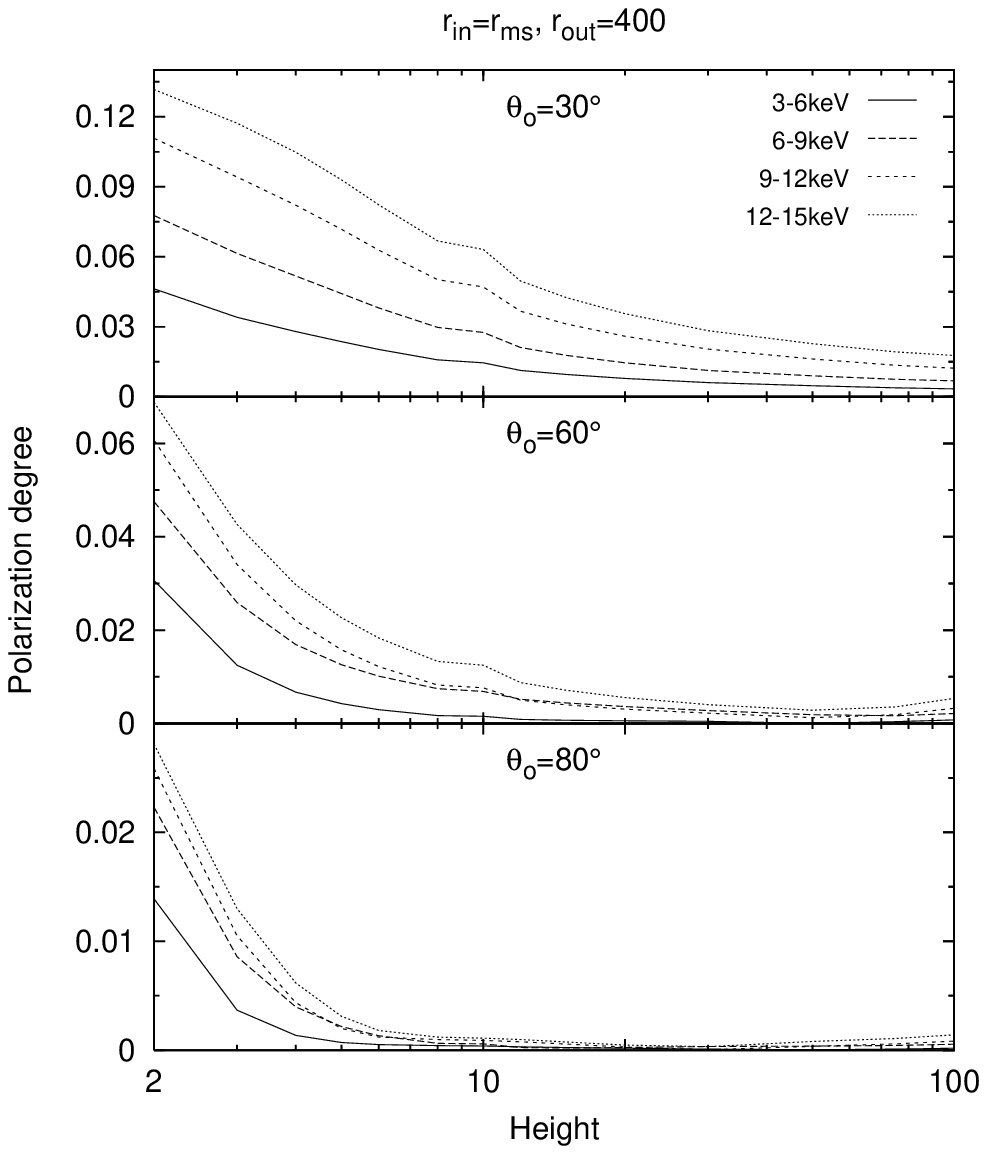}
\caption{Net polarization degree of the total (primary
plus reflected) signal as a function of $h$.
Left panel: $r_{\rm{}in}=6$; right panel: $r_{\rm{}in}=1.20$.
The curves are parametrized by the corresponding energy range.}
\label{poldeg1}
\end{figure*}

\begin{figure*}
\includegraphics[width=0.45\textwidth]{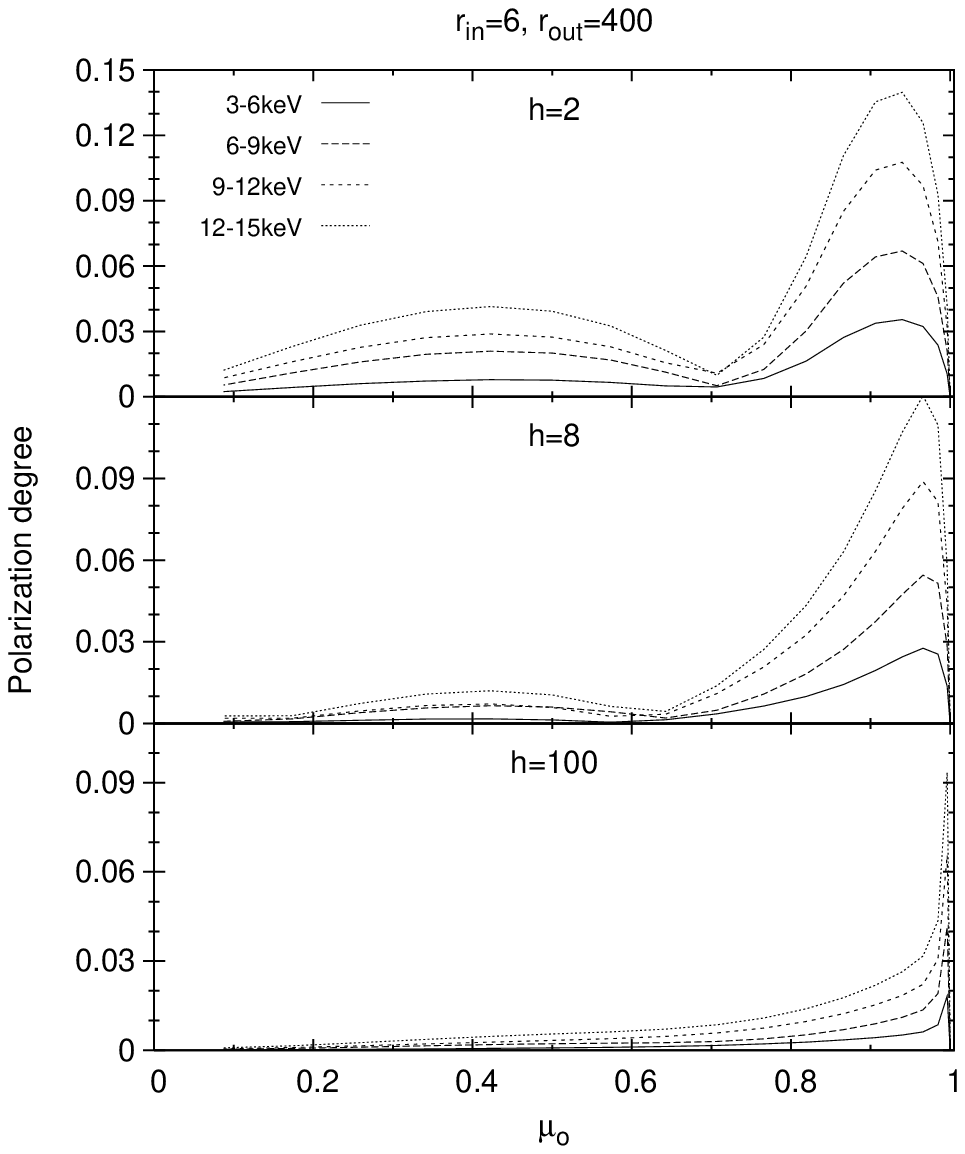}
\hfill
\includegraphics[width=0.45\textwidth]{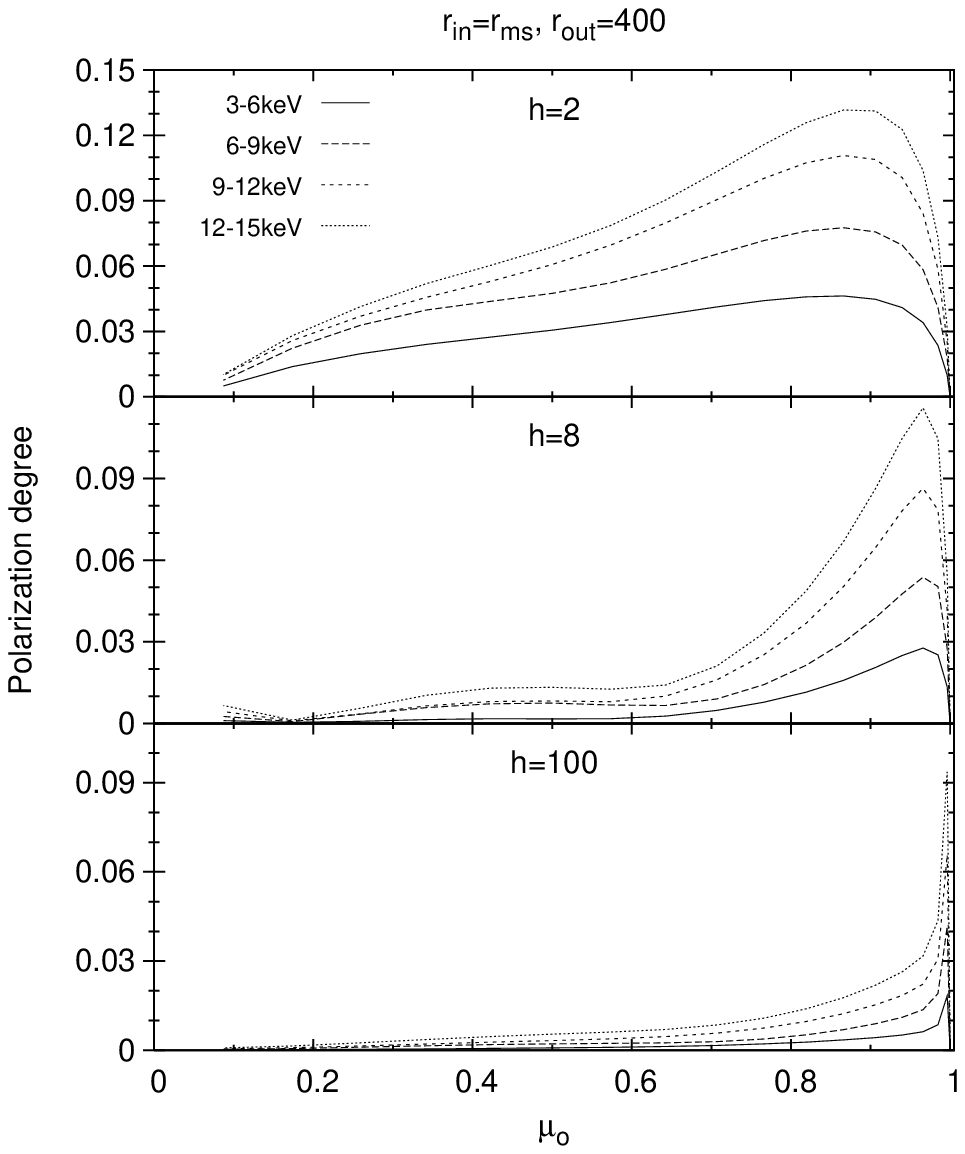}
\caption{Net polarization degree of the total (primary
plus reflected) signal as a function of $\mu_{\rm{}o}$.
The same model is shown as in the previous figure.}
\label{poldeg2}
\end{figure*}

\section{Results}
The reflection component has been computed by a Monte-Carlo code. The
number of reflected photons is proportional to the incident flux
$N_{\rm{}i}^{S}(E_{\rm{}p})$ arriving from the primary source,
\begin{equation}
N_{\rm{}i}^{S}(E_{\rm{}i})=N_{\rm{}p}^{\Omega}(E_{\rm{}p})\frac{{\rm{}d}
\Omega_{\rm{}p}}{{\rm{}d}S_{\rm{}loc}}\,,
\end{equation} 
where
$N_{\rm{}p}^{\Omega}(E_{\rm{}p})=N_{0{\rm{}p}}\,E_{\rm{}p}^{-\Gamma}$
represents an isotropic and steady power-law primary emission that is
emitted into the solid angle ${\rm{}d}\Omega_{\rm{}p}$ and eventually
illuminates the local area element ${\rm{}d}S_{\rm{}loc}$ on the disc
(see refs.\  \cite{dovciak04a,dovciak04b,dovciak04c} for a more 
detailed description of computations and for notation).
Four Stokes parameters, $I_{\nu}$, $Q_{\nu}$, $U_{\nu}$ and $V_{\nu}$,
entirely describe polarization properties of the scattered light. One
has to distinguish the quantities that are determined locally at the
point of emission on the disc surface (index `loc') and those relevant
to a distant observer (`o'). We introduce specific Stokes parameters,
\begin{equation}
i_\nu\equiv\frac{I_{\nu}}{E}\,,\quad q_\nu\equiv\frac{Q_{\nu}}{E}\,,\quad
u_\nu\equiv\frac{U_{\nu}}{E}\,,\quad v_\nu\equiv\frac{V_{\nu}}{E}\,,
\end{equation}
and then specific Stokes parameters per energy bin, i.e.\ $\Delta
i_{\rm{}o}$, $\Delta q_{\rm{}o}$, $\Delta u_{\rm{}o}$ and $\Delta
v_{\rm{}o}$. The latter quantities are directly measurable,
specifying the fluxes of photons with a given polarization.
One can write
\begin{eqnarray}
\label{S1}
{\Delta}i_{\rm{}o}(E,\Delta E) & = & N_0\int{\rm{}d}S\,\int{\rm{}d}E_{\rm{}loc}\,
i_{\rm{}loc}(E_{\rm{}loc})\,F\, ,\\
\label{S2}
{\Delta}q_{\rm{}o}(E,\Delta E) & = & N_0\int{\rm{}d}S\,\int{\rm{}d}E_{\rm{}loc}\,
\Big[q_{\rm{}loc}(E_{\rm{}loc})\cos{2\Psi} \nonumber \\
 && -u_{\rm{}loc}(E_{\rm{}loc})\sin{2\Psi}\Big]\,F\, ,\\
 \label{S3}
 {\Delta}u_{\rm{}o}(E,\Delta E) & = & N_0\int{\rm{}d}S\,\int{\rm{}d}E_{\rm{}loc}\,
 \Big[q_{\rm{}loc}(E_{\rm{}loc})\sin{2\Psi} \nonumber \\
 && +u_{\rm{}loc}(E_{\rm{}loc})\cos{2\Psi}\Big]\,F\, ,\\
  \label{S4}
  {\Delta}v_{\rm{}o}(E,\Delta E) & = & N_0\int{\rm{}d}S\,\int{\rm{}d}E_{\rm{}loc}\,
  v_{\rm{}loc}(E_{\rm{}loc})\,F\, ,
  \end{eqnarray}
where $F\equiv F(r,\varphi)=g^2\,l\,\mu_{\rm{}e}\,r$ is the transfer
function, $g$ being the total energy shift between observed and emitted
photons, $l$ the lensing effect, $\mu_{\rm{}e}$ the cosine of the
emission angle, and $\Psi$ the angle by which a vector rotates while it
is parallelly transported along the light geodesic.

In Figures \ref{pol} and \ref{pol1} we show, respectively, the energy
dependence  of polarization angle and polarization degree due to
reflected and reflected plus direct radiation. One can see that the
polarization of reflected radiation can be as high as thirty percent for
small inclinations and small heights of the primary source. Polarization
of the reflected radiation depends only weakly on energy, except
for the region close to the iron edge at $\sim7.2\,$keV.
In order to compute observable characteristics one has to
combine the primary power-law continuum with the reflected component.
The polarization
degree of the resulting signal depends on the ratio between the two
components and also on the energy range of an observation. The overall degree of
polarization increases with energy (see bottom panels in
Figs.~\ref{pol}--\ref{pol1}) due to the fact that the intensity of radiation
from the primary source diminishes, the intensity of the reflected
radiation increases with energy (in the energy range $3-15\,$keV) and
the polarization of the reflected light alone stays roughly constant.

In our computations we assumed that the irradiating source emits
isotropically and its light is affected only by gravitational redshift
and lensing, according to the source location at $z=h$ on axis. This
results in a dilution of primary light by factor
${\sim}g_{\rm{}h}^2(h,\theta_{\rm{}o})\,l_{\rm{}h}(h,\theta_{\rm{}o})$,
where $g_{\rm{}h}^2=1-2h/(a^2+h^2)$ is square of the redshift of primary
photons reaching directly the observer, $l_{\rm{}h}$ is the
corresponding lensing factor. Here, the redshift is the dominant
relativistic term, while lensing of primary photons is a few percent at
most and can be safely ignored. Anisotropy of primary radiation may
further attenuate or amplify the polarization degree of the final
signal, while the polarization angle is almost independent of this
influence as long as the primary light is itself unpolarized.

The polarization of scattered light is also shown in Fig.~\ref{poldeg},
where we plot the polarization degree and the change of the polarization
angle as functions of $h$. Notice that in the Newtonian case only
polarization angles of $0^{\circ}$ or $90^{\circ}$ would be expected for
reasons of symmetry. The change in angle is due to gravitation for which
we assumed a rapidly rotating black hole. The two panels of this figure
correspond to different locations of the inner disc edge:
$r_{\rm{}in}=6$ and $r_{\rm{}in}=1.20$, respectively. The curves are
strongly sensitive to $r_{\rm{}in}$ and $h$, while the dependence on
$r_{\rm{}out}$ is weak for a large disc (here $r_{\rm{}out}=400$).
Sensitivity to $r_{\rm{}in}$ is particularly appealing if one remembers
practical difficulties in estimating $r_{\rm{}in}$ by fitting spectra.
The effect is clearly visible even for $h\sim20$. Graphs corresponding
to $r_{\rm{}in}=6$ and $a=0.9987$ resemble the non-rotating case ($a=0$)
quite closely because dragging effects are most prominent near horizon.

Figure~\ref{polangle1} shows the polarization degree and angle as
functions of the observer's inclination. Again, by comparing the two
cases of different $r_{\rm{}in}$ one can clearly recognize that the
polarization is sensitive to details of the flow near the inner disc
boundary. Finally, dependence of the polarization degree of overall
radiation (primary plus reflected) on the height of the primary source
and the observer inclination in different energy ranges is shown in
Figures \ref{poldeg1}--\ref{poldeg2}.

\section{Conclusions}
We examined strong-gravity polarization features in X-rays reflected
from accretion discs. In order to compute directly observable
characteristics one has to combine the primary continuum with the
reflected component.
Polarization degree of the resulting signal depends on mutual proportion
of the two components and the energy range of observation. Polarization
properties represent the scattering mechanism, source geometry as well
as the gravitational field structure acting on reflected photons. New
generation photoelectric polarimeters in the focal plane of large area
optics, such as those foreseen for {\it{}Xeus}, can probe polarization
degree of the order of one percent in bright AGN, making polarimetry,
along with timing and spectroscopy, a tool for exploring the properties
of the accretion flows in the vicinity of black holes.

The authors gratefully acknowledge support from Czech Science Foundation
grants 205/03/0902 (VK) and 205/05/P525  (MD), and from the Grant Agency 
of the Academy of Sciences (IAA\,300030510). GM acknowledges financial
support from Agenzia Spaziale Italiana (ASI) and Ministero
dell'Istruzione, dell'Universit\`a e della Ricerca (MIUR), under grant 
{\sc cofin--03--02--23}.

\end{document}